\documentclass[twocolumn,aps]{revtex4}
\usepackage{epsfig}

\newcommand{\beq}{\begin{eqnarray}}
\newcommand{\eeq}{\end{eqnarray}}

\begin{document}

\title{Three-body structure of the $nn\Lambda$ system with  $\Lambda N-\Sigma N$
coupling}

\author{$^1$E. Hiyama}

\address{$^1$Nishina Center for
Accelerator-Based Science,
Institute for Physical and Chemical
Research (RIKEN), Wako
351-0198, Japan}

\author{$^{1,2}$S. Ohnishi}

\address{$^2$Department of Physics, Tokyo Institute of Technology,
Meguro 152-8551, Japan}

\author{B.\ F. Gibson}

\address{Theoretical Division, Los Alamos
National Laboratory. Los Alamos, New Mexico 87545, USA}

\author{Th.\ A. Rijken}

\address{IMAPP, Radboud University, Nijmegen, The Netherlands}

%

\begin{abstract}
The structure of the three-body $nn\Lambda$ system, which has been observed recently 
by the HypHI collaboration, is investigated taking $\Lambda N-\Sigma N$ coupling explicitly 
into account. The $YN$ and $NN$ interactions employed in this work  reproduce the
binding energies of $^3_{\Lambda}$H, $^4_{\Lambda}$H and $^4_{\Lambda}$He.
We do not find any $^3_{\Lambda}n$ bound state, which contradicts the interpretation
of the data reported by the HypHI collaboration. 
\end{abstract}

\maketitle

\section{Introduction}

In 2012 the HypHI Collaboration  \cite{Rappold2013} reported evidence that the 
neutron-rich  system $nn\Lambda$ was bound (the hypernucleus $^3_{\Lambda}n$), 
based upon observation of the two-body and three-body decay modes.
This claim is very significant for hypernuclear physics for the following reason:
One research goal in hypernuclear physics is to study new dynamical
features obtained by adding a $\Lambda$ particle to a nucleus.
In this vein it is interesting to explore the resulting structure of neutron-rich
$\Lambda$ hypernuclei.
Core nuclei corresponding to neutron-rich systems can be
weakly bound halo states or even resonant states.
When a $\Lambda$ particle is added to such a nuclear core,
the resultant hypernuclei will become more stable against neutron decay,
due to the attraction of the $\Lambda N$ interaction.  The $^3_{\Lambda}n$
system would be the lightest such example that might be bound, one in which a 
$\Lambda$ is bound to a di-neutron ($nn$) pair.

Another important goal of studying $\Lambda$ hypernuclei is the extraction 
of information about the effect of $\Lambda N-\Sigma N$ coupling.  
For this purpose many authors have performed few-body calculations:
Miyagawa {\it et al.}~\cite{Miyagawa1995}, performed Faddeev calculations 
for  $^3_{\Lambda}$H using realistic $YN$ interactions -- the  Nijmegen 
soft core 89 (NSC89) \cite{Rijken89} and the Juelich potential  \cite{Julich};  
these authors confirmed that  $\Lambda N-\Sigma N$ coupling
plays a crucial role in obtaining a bound state of $^3_{\Lambda}$H.
To further investigate $\Lambda N-\Sigma N$ coupling, $^4_{\Lambda}$He and
$^4_{\Lambda}$H are perhaps most useful because both of the spin-doublet 
states have been observed.  To study this feature in the $A=4$ hypernuclei,
the authors of Ref.~\cite{Gibson1972} utilized a coupled-channel two-body 
model of $^3{\rm He}(3{\rm H})+\Lambda/\Sigma$, 
and later, Akaishi {\it et al.}~\cite{Akaishi2000} analyzed the role of the
$\Lambda N-\Sigma N$ coupling for the $0^+$-$1^+$ splitting within
the same framework.  It is necessary to perform four-body 
coupled-channel calculations to investigate the role of $\Lambda N-\Sigma N$ 
coupling.  Four-body coupled-channel calculations with separable potentials 
that were central in nature were performed by the authors of Ref.~\cite{Gibson1988}.
Carlson then carried out four-body calculations with the NSC89 potential model 
using variational Monte Carlo methods \cite{Carlson1991}, and he
obtained binding energies  with statistical errors of 100 keV. 
Later, one of the authors (E.H.) performed four-body coupled-channel 
calculations~\cite{Hiyama2001} using a $\Lambda N-\Sigma N$ 
coupled-channel $YN$ potential~\cite{Shinmura2001} with central, 
spin-orbit, and tensor terms which simulates the scattering 
phase shifts given by NSC97f~\cite{RSY99}. 
A four-body calculation of $^4_{\Lambda}$He and $^4_{\Lambda}$H and a
five-body calculation of $^5_{\Lambda}$He using the same $YN$ interaction
were performed by Nemura {\it et al.}~\cite{Nemura2002}.
 More sophisticated four-body calculations of  $^4_{\Lambda}$He and 
 $^4_{\Lambda}$H using realistic $NN$ and $YN$ interactions were 
 performed by Nogga {\it et al.}~\cite{Nogga2002}.  In the A=4 system 
 $\Lambda N-\Sigma N$ coupling plays a crucial role in charge symmetry 
 breaking as well as in the ground states being 0$^+$.  However, the nature
 of $\Lambda N-\Sigma N$ mixing is not fully understood.  The
 hypertriton $^3_{\Lambda}$H is bound.  If $^3_{\Lambda}n$ were also 
 bound, the pair would provide complementary information about 
 $\Lambda N-\Sigma N$ coupling.
 
It is thought that $\Lambda N-\Sigma N$ coupling may also play an 
important role in the structure of heavier neutron-rich $\Lambda$ hypernuclei, 
because of the increasing total isospin.  For example, a recent FINUDA 
experiment \cite{FINUDA} reported a heretofore unobserved bound state of the 
superheavy hydrogen-$\Lambda$ hypernucleus $^6_{\Lambda}$H.
Furthermore, in 2013, another light neutron-rich $\Lambda$ hypernucleus
bound state, $^7_{\Lambda}$He, was observed at JLAB \cite{JLAB}.
Among the observed neutron-rich $\Lambda$ hypernuclei,
$^3_{\Lambda}n$ would be a unique neutron-rich $\Lambda$ hypernucleus,
one containing no protons.  Thus, measuring the binding energy of a bound 
$^3_{\Lambda}n$ system would contribute directly to understanding the 
structure of neutron-rich $\Lambda$ hypernuclei and to understanding the 
nature of $\Lambda N-\Sigma N$ coupling.  However, no binding energy 
was reported for $^3_{\Lambda}n$ by the HypHI Collaboration.

Given the current situation, an important theoretical issue to address is whether 
a bound state of $^3_{\Lambda}n$ can exist.  In addition, it is imperative that 
an estimate of the binding of any such bound system be calculated.  In such 
an analysis it is important that the interactions employed be realistic and 
constrained by the known data for the s-shell $\Lambda$ hypernuclei.  
$\Lambda N-\Sigma N$ coupling should be a component of the $YN$ 
interactions used in the investigation.  Previously, in Ref.~\cite{Downs1959}, 
a three-body calculation of  $^3_{\Lambda}n$ was performed using a 
variational method and a $\Lambda N$ interaction without $\Lambda N-\Sigma N$ 
coupling; it was concluded that no bound state existed.  Recently, a Faddeev 
calculation for this system was performed taking $\Lambda N-\Sigma N$ 
coupling explicitly into account \cite{Garcilazo2007}; again, no bound state 
was found.  Thus, previous theoretical predictions are inconsistent with the 
conclusion based upon the new observation of the decay modes for a 
bound $^3_{\Lambda}n$ as reported by the HYPHI collaboration~\cite{Rappold2013}.  
Here, motivated by experimental and theoretical studies, the goal of 
this work is to investigate the possibility of the existence of a bound state in 
this system using realistic $YN$ interactions with $\Lambda N-\Sigma N$ 
coupling explicitly taken into account.  We use the  Gaussian Expansion Method 
in our calculations.  It is important that we employ a $YN$ interaction which 
reproduces the binding energies of the observed $s$-shell $\Lambda$ 
hypernuclei such as $^3_{\Lambda}$H, $^4_{\Lambda}$H and 
$^4_{\Lambda}$He in addition to a realistic $NN$ interaction.  For this purpose, 
we employ the NSC97f simulated $YN$ interaction~\cite{Shinmura2001} 
which was used in Refs.~\cite{Hiyama2001,Nemura2002}, because this 
interaction reproduces reasonably the binding energies of those $\Lambda$ 
hypernuclei in combination with the AV8 $NN$ interaction.  

This article is organized as follows:
In Sec II., the method and interactions used in the three-body calculation for
the $nn\Lambda$ system are described. The numerical results and a
corresponding discussion are presented in Sec.  III.  A summary is given in 
Sec. IV.

\section{Method and Interaction}

\begin{figure}[htb]
\begin{center}
\epsfig{file=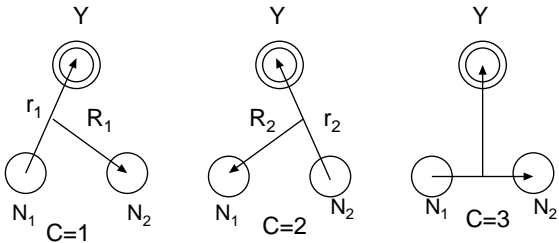,scale=0.45}
\end{center}
\caption{Jacobi coordinates 
of the $nn \Lambda$ three-body 
system. Antisymmetrization 
of the two neutrons is to be made.
}
\label{fig:jacobi}
\end{figure}

The total three-body wavefunctions for $^3_{\Lambda}$H and $^3_{\Lambda}n$ 
are described as a sum
of amplitudes for all rearrangement channels ($c=1 -3$)
of Fig.~\ref{fig:jacobi} in the $LS$ coupling scheme:
\begin{eqnarray}
&&\Psi_{JM}(^3_{\Lambda}{\rm He},\,\, ^3_{\Lambda}n) =  \sum_{Y=\Lambda,\Sigma}
\sum_{c=1}^3 \sum_{n\ell, NL, I} \sum_{sS, tt_Y}  {\cal A}  \nonumber \\
&& \times \Bigg[
 \Big[ \, 
  \Big[\, [ \phi^{(c)}_{nl}({\bf r}_c)\,
         \psi^{(c)}_{NL}({\bf R}_c) ]_I 
               \, \Big[\, [\chi_{\frac{1}{2}}(N_1)
              \chi_{\frac{1}{2}}(N_2)]_s \,
       \chi_{\frac{1}{2}}(Y) \Big]_S \Big]_{JM} \,\, 
\nonumber \\
\,\,&& \quad \times\,  \Big[\, [\eta_{\frac{1}{2}}(N_1)
              \eta_{\frac{1}{2}}(N_2)]_t \,
       \eta_{t_Y}(Y) \Big]_{T}  \Bigg]
\end{eqnarray}

Here, $\cal{A}$ is the two-nucleon antisymmetrization operator and the 
$\chi$'s and $\eta$'s are the spin and isospin functions, respectively,
with the isospin $t_{Y=0 (1)}$ for $Y=\Lambda (\Sigma)$.
$T$ is total isospin, 0 for $^3_{\Lambda}$H and 1 for $^3_{\Lambda}n$.
$J$ is the total spin, $1/2^+$, for both hypernuclei.
In addition, to investigate the contribution of the $^3S_1$ state in the 
$YN$ interaction, we calculate the binding energy for the $J=3/2^+$ of 
$^3_{\Lambda}$H. The functional form of $ \phi^{(c)}_{nl}({\bf r}_c)$ is
taken as  $\phi^{(c)}_{nl}({\bf r}_c)=r^\ell e^{{{-r/r_n}}^2} Y_{\ell m}(\hat{\bf r})$,
where the Gaussian range parameters are chosen to satisfy a geometrical
progression ($r_n=r_1a^{n-1}; n=1 \sim n_{\rm max}$),
and similarly for $\phi_{NL}(\bf R)$.   Three basis functions were verified to be 
sufficient for describing both the short-range correlations and the long-range 
tail behavior of the few-body systems \cite{Kamimura1988, Kameyama1989,Hiyama2003}.

The details of the four-body calculation,  $^4_{\Lambda}$H and $^4_{\Lambda}$He,
can be found in Ref.\cite{Hiyama2001}.

The $YN$ interaction employed in the three- and four-body systems  is the same 
as in Refs.\cite{Hiyama2001,Nemura2002}.  Namely, the $YN$ potential simulates 
the scattering phase shifts given by NSC97f.
The $YN$ interaction is represented as

\begin{eqnarray}
^{2S+1}V_{NY{\rm -}NY'}(r) & = & \sum_i (^{2S+1}V^C_{NY{\rm -}NY'} {\rm e}^{-(r/\beta_i)^2} \nonumber \\
& + &^{2S+1}V^{T}_{NY{\rm -}NY'}S_{12}{\rm e}^{-(r/\beta_i)^2}   \nonumber \\
&+ &^{2S+1}V_{NY {\rm -}NY'}^{LS}{\bf LS}\, {\rm e}^{-(r/\beta_i)^2} )
 \nonumber \\  &{\rm for }&  \quad  T=1/2 \,\, ,    \nonumber \\
^{2S+1}U_{N\Sigma{\rm -}N\Sigma} & = & \sum_i ( ^{2S+1}U^C_{N\Sigma{\rm -}N\Sigma}\,{\rm e}^{-(r/\beta_i)^2} \nonumber \\
& + & ^{2S+1}U^{T}_{N\Sigma{\rm -}N\Sigma}S_{12} {\rm e}^{-(r/\beta_i)^2}   \nonumber \\
&+ &^{2S+1}U_{N\Sigma{\rm -}N\Sigma}^{LS}{\bf LS}\, {\rm e}^{-(r/\beta_i)^2})  
 \nonumber \\
 & {\rm for }&  \quad T=3/2 \,\,  ,   
\label{eq:potential}
\end{eqnarray}
with $Y,Y'=\Lambda$ or $\Sigma$. Here, $C, T, LS$ mean central, tensor and spin-orbit terms
with two-range Gaussian forms. 
The potential parameters are listed in Table~\ref{tab:YN}.
\begin{table}[h]
\caption{Parameters of the $YN$ interaction defined in
Eq. (\ref{eq:potential}). Range parameters are in fm and strengths are in MeV.}
\vskip 0.2cm
 \begin{tabular}{ccc}
 \hline
 \hline
$i$   & 1 &2    \\  
$\beta_i$  &0.5   &1.2 \\
\hline
\vspace {-4 mm} \\
$^1V^C_{N\Lambda {\rm -} N\Lambda}$   &$\enskip732.08$ &$-99.494$   \\
$^1V^C_{N\Lambda {\rm -} N\Sigma}$ &$\enskip 61.223$    &$-15.977$  \\
$^1V^C_{N\Sigma {\rm -} N\Sigma}$  &\enskip 1708.0    &$\enskip 80.763$ \\
$^1U^C_{N\Sigma {\rm -} N\Sigma}$  &\enskip 695.39    &$-109.37$ \\
$^3V^C_{N\Lambda {\rm -} N\Lambda}$  &\enskip 1068.8    &$-45.490$ \\
$^3V^C_{N\Lambda {\rm -} N\Sigma}$ &$-770.21$    &$\enskip 68.274$  \\
$^3V^C_{N\Sigma {\rm -} N\Sigma}$  &\enskip 863.76    &$\enskip 28.284$ \\
$^3U^C_{N\Sigma {\rm -} N\Sigma}$  &$-181.08$    &$\enskip 23.282$ \\
$^3V^T_{N\Lambda {\rm -} N\Lambda}$  &$-243.31$    &$-10.413$ \\
$^3V^T_{N\Lambda {\rm -} N\Sigma}$  &$\enskip 287.54$    &$\enskip 62.438$ \\
$^3U^T_{N\Sigma {\rm -} N\Sigma}$  &$\enskip 333.05$    &$\enskip 22.234$ \\
$^3V^{LS}_{N\Lambda {\rm -} N\Lambda}$  &$\enskip 1023.8$    &$-17.195$ \\
$^3V^{LS}_{N\Lambda {\rm -} N\Sigma}$  &$-19.930$    &$\enskip 22.299$ \\
$^3U^{LS}_{N\Sigma {\rm -} N\Sigma}$  &$-462.31$    &$\enskip 0.0023$ \\
\vspace {-3 mm} \\
 \hline
 \end{tabular}
 \label{tab:YN}
\end{table}

The interaction reproduces the observed binding energy of $^3_{\Lambda}$H:
the calculated $\Lambda$ binding energy, $B_{\Lambda}$, 0.19 MeV is
consistent with the observed data [$B_{\Lambda} (^3_{\Lambda}$H =$0.13 \pm 0.05$ MeV].
Furthermore, the calculated energies  of $^4_{\Lambda}$H and
$^4_{\Lambda}$He are 2.33 MeV and 2.28 MeV, respectively.
For the $NN$ interaction we employ the AV8 potential \cite{Pudliner1997}.

\section{Results and Discussion}

\begin{figure}[htb]
\begin{center}
\epsfig{file=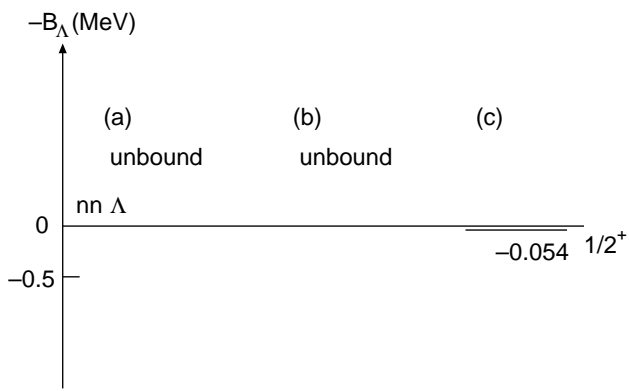,scale=0.45}
\caption{Calculated $\Lambda$-separation energy for $^3_{\Lambda}n$
with (a) $^3V_{N\Lambda -N\Sigma }^T\times 1.00$, 
(b) $^3V_{N \Lambda -.N\Sigma}^T\times 1.10$, and (c) $^3V_{N \Lambda-N\Sigma}^T \times 1.20$.
The energy is measured with respect to the $nn\Lambda$ three-body breakup threshold.}
\label{fig:nnl}
\end{center}
\end{figure}
\begin{figure}[htb]
\begin{center}
\epsfig{file=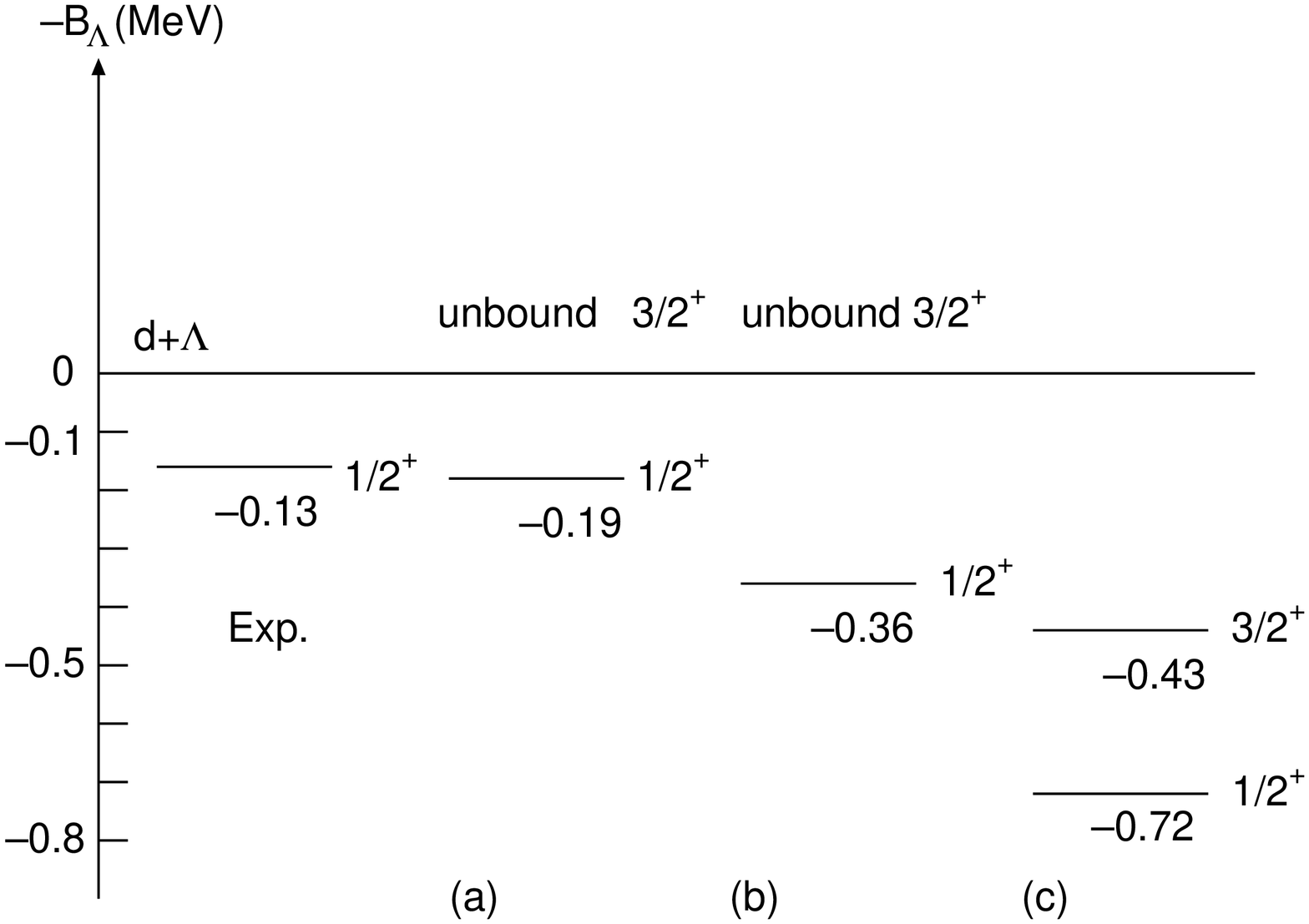,scale=0.40}
\caption{Calculated $\Lambda$-separation energy for $^3_{\Lambda}$H
with (a) $^3V_{N\Lambda -N\Sigma }^T\times 1.00$, 
(b) $^3V_{N \Lambda -.N\Sigma}^T\times 1.10$, and (c) $^3V_{N \Lambda-N\Sigma}^T \times 1.20$.
The energy is measured with respect to the $np\Lambda$ three-body breakup threshold.}
\label{fig:h3l}
\end{center}
\end{figure}
\begin{figure}[htb]
\begin{center}
\epsfig{file=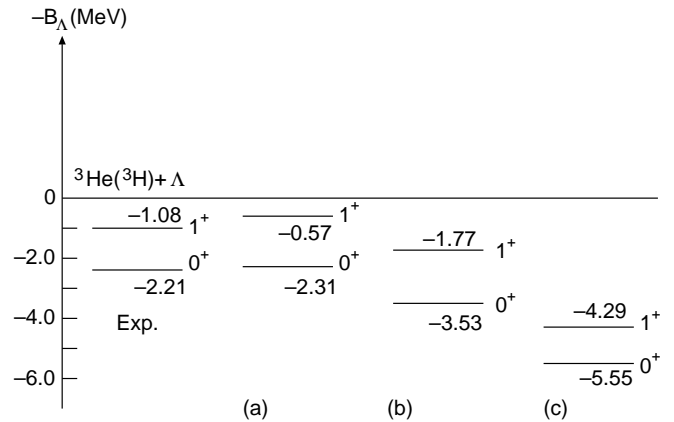,scale=0.40}
\caption{Calculated $\Lambda$-separation energy of ground state in $^4_{\Lambda}$He
with (a) $^3V_{N\Lambda -N\Sigma }^T\times 1.00$, 
(b) $^3V_{N \Lambda -.N\Sigma}^T\times 1.10$, and (c) $^3V_{N \Lambda-N\Sigma}^T \times 1.20$.
The energy is measured with respect to the $^3{\rm He}+\Lambda$ breakup threshold.
In parenthesis is the energy of $^4_{\Lambda}$H.}
\label{fig:h4l}
\end{center}
\end{figure}

Before discussing results for the $^3_{\Lambda}$n system, 
 we consider  two possibilities: (1) We investigate first the  possibility of having a bound state in
the $nn\Lambda$ system by tuning the $YN$ potential while maintaining consistency with
the binding energies of $^3_{\Lambda}$H and $^4_{\Lambda}$H and $^4_{\Lambda}$He.
(2) We investigate second the  possibility of having a bound state in the $nn\Lambda$ system 
by tuning the $nn$ $^1S_0$ state while maintaining consistency with
the binding energies of $^3$H.

First, let us consider case (1).
In $^3_{\Lambda}n$, the $nn$ pair is in the spin-singlet state ($s=0$, spin antiparallel),
while in the $^3_{\Lambda}$H system, the $np$ pair is in the spin-triplet state ($s=1$, 
spin parallel).  The difference in the spin value of $[NN]_{s=1 {\rm or} 0}$ leads to 
different contributions of the $\Lambda N$ spin-spin interaction to the
doublet splitting.
In the $^3_{\Lambda}n$ system, the $1/2^+$ ground state includes $YN$ spin-singlet and 
spin-triplet states, while in the $^3_{\Lambda}$H system, the $1/2^+$ ground state is 
dominated by the $YN$ spin-singlet state.
To investigate the contribution of the spin-singlet state and spin-triplet state, in Table~\ref{tab:Gal}(a),
we list expectation values of  the $S=0$ and $S=1$ states of $V_{\Lambda N-\Lambda N}$,
$V_{\Lambda N-\Sigma N}$ and $V_{\Sigma N-\Sigma N}$.
We find that the $S=0$ state in the $V_{\Lambda N-\Lambda N}$ term dominates in 
the binding energy of $^3_{\Lambda}$H.  Also, it is found that the
contribution of $V_{\Lambda N-\Sigma N}$ coupling  in the $S=1$ state  to the
binding energy is large. For the $S=1$ state
the calculated expectation value of the tensor component in $V_{\Lambda N-\Sigma N}$
is $-0.47$ MeV. This means that the $S=1$ state of the central  $V_{\Lambda N-\Sigma N}$
component  dominates in the binding energy of $^3_{\Lambda}$H. 
Therefore, we tune the spin-triplet state of the $YN$ interaction in a manner that does not
affect the binding energy of $^3_{\Lambda}$H significantly.
To accomplish this, we multiply the strength of the tensor part of the 
$\Lambda N-\Sigma N$ coupling by a factor,
because the tensor part of the $\Lambda N-\Sigma N$ coupling acts in the 
spin-triplet state of  the $YN$ interaction.
To verify the consistency of the effect due to the spin-triplet state of the $YN$ interaction,
we also investigated the binding energy of the $J=3/2^+$ state in $^3_{\Lambda}$H, which
has not been observed experimentally.

The calculated energy of $^3_{\Lambda}$n with $T=1$ and $J^\pi =1/2^+$ is
illustrated in Fig.~\ref{fig:nnl}.  In Fig.~\ref{fig:h3l} we show the binding energies of the $J=1/2^+$ 
and $3/2^+$ states in $^3_{\Lambda}$H. 
The energy in Fig.~\ref{fig:nnl}(a) is obtained using the $YN$ interaction which reproduces the 
binding energy of the ground state in $^3_{\Lambda}$H, the $J=1/2^+$ state.
In Fig.~\ref{fig:nnl}(b) and \ref{fig:nnl}(c), we multiply the tensor part of the 
$\Lambda N-\Sigma N$ coupling by the factor 1.10 and 1.20.
In Fig.~\ref{fig:nnl}(b), we still obtain no bound state in the $nn\Lambda$ system.
When $^3V^T_{N\Lambda {\rm -} N\Sigma}$ is multiplied by 1.20,
then we obtain a very weakly  bound state ($-0.054$ MeV) with respect to the 
$nn\Lambda$ three-body breakup threshold. 
In order to judge whether the adjusted  $^3V^T_{N\Lambda {\rm -} N\Sigma}$
is reasonable, we calculated the binding energies of  the ground and the excited states 
in $^3_{\Lambda}$H as shown in Fig.~\ref{fig:h3l}.
In  Fig.~\ref{fig:h3l}(a) the binding energy of the ground state, $J=1/2^+$,  
is  in good agreement with the observed data. 
As one should anticipate, we find the energy of $^3_{\Lambda}$H becomes deeper
with increasing strength of  $^3V^T_{N\Lambda {\rm -} N\Sigma}$.
When the strength of  $^3V^T_{N\Lambda {\rm -} N\Sigma}$is multiplied by the factor 1.20,
which leads to a bound state in the $nn\Lambda$ system,
the energy of $^3_{\Lambda}$H is over bound
($-0.7$ MeV) compared with the observed data ($-0.13$ MeV).
In addition, we find a bound $J=3/2^+$ excited state ($-0.4$ MeV) of
$^3_{\Lambda}$H, for which there is no experimental evidence.
To make clear the contribution of the $YN$ potential in the binding energies of 
$^3_{\Lambda}n$ and $^3_{\Lambda}$H, we show 
the expectation values of the $YN$ potential in Table~\ref{tab:Gal}(b) and \ref{tab:Gal}(c).
We find that, with the tensor term of $V_{\Lambda N-\Sigma N}$ 
enhanced  by 20 \%, the corresponding contribution to the potential expectation 
values is much larger.

\begin{table*}
\caption{Expectation values of  the $YN$ interaction for (a) $J=1/2^+$ state, 
(b) $J=3/2^+$ state of $^3_{\Lambda}$H, and (c) $J=1/2^+$  state of $^3_{\Lambda}n$.
In the case of (a), the calculated $B_{\Lambda}=0.19$ MeV.
In (b), the calculated $B_{\Lambda}=0.43$ MeV using
 $^3V^T_{N\Lambda {\rm -} N\Sigma} \times 1.20$. 
In (c), the calculated $B_{\Lambda}=0.054$ MeV using
 $^3V^T_{N\Lambda {\rm -} N\Sigma} \times 1.20$.} 
\vskip 0.2cm
 \begin{tabular}{ccccc}

(a) $J=1/2^+$ for $^3_{\Lambda}$H \\
  & $<V_{\Lambda N {\rm -} \Lambda N}>$
 &  $<V_{\Lambda N {\rm -} \Sigma N}>$
 & $<V_{\Sigma N {\rm -} \Sigma N}>$  &$<V_{YN}>$ \\
\vspace {-4 mm} \\
\hline
\vspace {-4 mm} \\
$S=0$ &$-2.12$  &$-0.04$   
 &\enskip 0.02   &$-2.14$  \\
$S=1$ &\enskip $0.04$   &$-1.49$  &$-0.03$   &$-1.48$  \\
 all   &$-2.08$
&$-1.53$  &$-0.01$   &$-3.62$  \\
\vspace {-3 mm} \\
 \hline \\
(b) $J=3/2^+$ for $^3_{\Lambda}$H \\
 & $<V_{\Lambda N {\rm -} \Lambda N}>$
 &  $<V_{\Lambda N {\rm -} \Sigma N}>$
 & $<V_{\Sigma N {\rm -} \Sigma N}>$  &$<V_{YN}>$ \\
\vspace {-4 mm} \\
\hline
\vspace {-4 mm} \\
$S=0$ &$-0.11$  &\enskip $0.02$   
 &\enskip 0.08   &$-0.01$  \\
$S=1$ &\enskip $7.87$   &$-34.69$  &\enskip $2.29$   &$-24.53$  \\
 all   &\enskip $7.76$
&$-34.67$  &\enskip $2.37$   &$-24.54$  \\
\vspace {-3 mm} \\
 \hline \\
(c) $J=1/2^+$ for $^3_{\Lambda}n$ \\
 & $<V_{\Lambda N {\rm -} \Lambda N}>$
 &  $<V_{\Lambda N {\rm -} \Sigma N}>$
 & $<V_{\Sigma N {\rm -} \Sigma N}>$  &$<V_{YN}>$ \\
\vspace {-4 mm} \\
\hline
\vspace {-4 mm} \\
$S=0$ &$-0.65$  &$-0.01$   
 &0.00   &$-0.66$  \\
$S=1$ &\enskip 2.45   &$-11.71$  &0.45   &$-8.81$  \\
 all   &\enskip $1.80$
&$-11.72$  &$0.45$   &$-9.47$  \\
\vspace {-3 mm} \\
 \hline 
\end{tabular}
\label{tab:Gal}
\end{table*}

To further investigate the reliability of the employed $YN$ interaction,
we calculate the binding energies of $^4_{\Lambda}$H and $^4_{\Lambda}$He.
In these two $\Lambda$ hypernuclei, we also see evidence of the important 
effect of charge symmetry breaking (CSB) in the $\Lambda N$ interaction.
The CSB effects appear in the ground state and excited state differences
 $\Delta_{\rm CSB}=B_{\Lambda}(^4_{\Lambda}{\rm He})-
B_{\Lambda}(^4_{\Lambda}{\rm H})$, the experimental values of
which are $0.35 \pm 0.06$ and $0.24 \pm 0.06$ MeV, respectively.
The $\Lambda$-separation energies of the ground states in $^4_{\Lambda}$He 
and $^4_{\Lambda}$H using the present $YN$ interaction are 2.28 MeV and 
2.33 MeV, respectively, which do not reproduce the CSB effect.
[To investigate CSB  in greater detail, it is planned to measure the 
$\Lambda$-separation energy of $^4_{\Lambda}$H at Mainz and Jlab.]
Here, it is not our purpose to explore the  CSB effect in $A=4$
$\Lambda$ hypernuclei.  Therefore, we adopt the average value of these 
hypernuclei.  That is,  as experimental data, we adopt
$B_{\Lambda}=2.21$ MeV and 1.08 MeV for the ground state and the
excited state, respectively.

In Fig.~\ref{fig:h4l} we illustrate the average binding energies of the $A=4$ hypernuclei.
In the case of Fig.~\ref{fig:h4l}(a), the calculated ground state energy reproduces the
data nicely, while the excited state is less bound than the observed data.
Then, as was done in the case of $^3_{\Lambda}$H, we tuned
  $^3V^C_{N\Lambda {\rm -} N\Sigma}$. 
As shown in Fig.~\ref{fig:h4l}(b) and Fig.~\ref{fig:h4l}(c), increasing the strength of 
 $^3V^C_{N\Lambda {\rm -} N\Sigma}$, means that both the $0^+$ and $1^+$ 
 states become overbound by 1-3 MeV. 
 
In addition, we adjusted other parts of the $YN$ potential such as 
$^3V^C_{N\Lambda {\rm -} N\Lambda}$, $^3V^C_{N\Lambda {\rm -} N\Sigma}$, {\it etc.}
However,  we could not find any modification of the $YN$ potential that produces a bound 
state in $^3_{\Lambda}n$  while maintaining consistency with the binding energies of A=3 
and 4 $\Lambda$ hypernuclei.

Next, we investigate case (2) the  possibility to have a bound state in $nn\Lambda$ 
by tuning the strength of the $nn$ $T=1, ^1S_0$ interaction.  It has been suggested 
that IF this channel has bound state, {\it i.e.,} a di-neutron state,  then it may be possible 
to describe the anomalies in neutron-deuteron elastic  scattering and the deuteron 
breakup reaction above threshold \cite{Witala2012}.  [However, we note that $pp$ 
scattering is well described by standard methods which do not admit a bound di-proton.  
Thus, the hypothesis of a bound di-neutron suggests strong CSB in the NN spin-singlet
interaction.]  This $nn$ spin-singlet channel does not contribute to the binding energy 
of $^3_{\Lambda}$H since the spin of the core nucleus, the deuteron, has spin 1.
On the other hand, the  
$^1S_0$ state of the $nn$ pair contributes to the binding  energy of $^3_{\Lambda}n$.
It also contributes to the energy of $^3$H. The observed energies of $^3$H and $^3$He
are -8.48 MeV and -7.72 MeV.
Then, it is interesting to ask what is the energy of $nn\Lambda$ as one tunes the 
 strength of the $nn$ $T=1, ^1S_0$ together with the predicted binding energy of $^3$H.

In Table~\ref{tab:energy}, we illustrate the component of  the $^1S_0$ state when multiplied 
by the factor $x$, scattering length, effective range, energy of di-neutron and energy of 
$^3_{\Lambda}n$.  When we use a $^1S_0$ component multiplied by the factor 1.13 
and 1.15, we have a bound state in the di-neutron system. However,
the $nn$ interaction multiplied by a factor of 1.13 does not produce any bound
state in $^3_{\Lambda}n$.
We find that the $^1S_0$ component, when multiplied by a factor of 1.35, leads to a 
bound state in $^3_{\Lambda}n$.
However, we see that as we increase the factor $x$ in
the $^1S_0$ component, that is, to 1.13 and 1.35, which produces a di-neutron 
bound state, the energy of $^3$H is overbound compared with the observed data. 
Then,  we do not have a bound state in  $^3_{\Lambda}n$ while maintaining
consistency with the observed data for $^3$H, {\it unless} we introduce a large
repulsive $nnp$ three-body force.  It is known that one needs a small (about 0.7 MeV)
attractive $npp$ three-body force to obtain agreement with the $^3$He binding
energy; our model value for the $^3$He binding energy is -7.12 MeV.  Thus, the 
hypothesis of a bound di-neutron would require a very large CSB in the $NNN$ 
three-body force, which is not easily understood.

\begin{table*}
\caption{Calculated binding energies of $nn\Lambda$, $E_{^3_{\Lambda}n}$,
 in the case that the $^1S_0$ component is
multiplied by the factor $x$.  The scattering length, $a_{nn}$,
effective range, $r_{\rm eff}$,  energies of di-neutron system, $\epsilon_{nn}$, 
 $^3$H, $E_{^3{\rm H}}$ are also listed for each $x$ factor.}
\label{tab:energy}
\begin{tabular}{cccccccc}
\hline   \hline
 &$x$   &$a_{nn}$(fm)  &$r_{\rm eff}$(fm)  &$\epsilon_{nn}$(MeV)
  &$E_{^3{\rm H}}$(MeV)   &$E_{^3_{\Lambda}n}$(MeV)  \\
\hline
&1.0  &$-23.7$  &2.78   &unbound  &$-7.77$     &unbound  \\
&1.13   &25.1  &2.40    &$-0.066$        &$-9.75$   &unbound   \\
&1.35   &6.88   &1.96    &$-1.269$        &$-13.93$    &$-1.272$   \\
\hline \hline
\end{tabular}
\end{table*}

\section{Summary}
Motivated by the reported observation of data suggesting a bound $^3_{\Lambda}n$ by
the HypHI collaboration, we have calculated the binding energy of this hypernucleus 
taking into account $\Lambda N-\Sigma N$ explicitly.  We consider it important to 
reproduce the observed data for $^3_{\Lambda}$H, $^4_{\Lambda}$H and 
$^4_{\Lambda}$He, and to be consistent with the energies of $^3$H and $^3$He.
For this purpose, we used simulated NSC97f $YN$ and AV8 $NN$ potentials which 
maintain consistency with the data mentioned above.
However, we did not find any bound state in the $^3_{\Lambda}n$ system.
Then, we investigated the possibility to produce such a bound state in $^3_{\Lambda}n$
(i) by tuning the strength of the $YN$ NSC97f potential and (ii) by tuning the $nn$ 
component of the $^1S_0$ potential.  When the strengths of the $YN$ NSC97f potential 
and the $nn$ $^1S_0$ potential are multiplied by 1.2, and 1.15, respectively, we 
obtain a very weakly bound state in the $^3_{\Lambda}n$ system.
However, in both cases, the calculated binding energies of the $s$-shell $\Lambda$ 
hypenuclei, $^3_{\Lambda}$H, $^4_{\Lambda}$H and $^4_{\Lambda}$He, and the
$s$-shell nucleus, $^3$H are overbound by   $0.6 \sim 3$ MeV in comparison
with the observed data.
That is,  we did not find any possibility to have a bound state in $^3_{\Lambda}n$.
However, the HypHI collaboration reported evidence for a bound state in this system;
such a finding is inconsistent with the present result. 
In order to corroborate the HypHI result, we should consider additional missing 
elements in the present calculation. 
Unfortunately, the HypHI data provide information on the life time of
this system  but  no binding energy.
If the experimentalists can provide a $^3_{\Lambda}n$ binding energy, it 
would be very helpful in explicating the mechanism that would produce such a bound 
state.  Further experimental study is urgently needed.

\parindent 15 pt

\section*{Acknowledgments}
The authors thank Dr. T. Saito, Dr. C. Rappold, and Prof.
Gal  for valuable discussions.
This work was supported by a Grant-in-Aid for Scientific Research from 
Monbukagakusho of Japan.  The numerical calculations were performed 
on the HITACHI SR16000 at KEK and YITP. This work was partly 
supported by JSPS Grant No. 23224006 and by RIKEN iTHES Project.
The work of BFG was performed under the auspices of the National 
Nuclear Security Administration of the U.S. Department of Energy at Los 
Alamos National Laboratory under Contract No DE-AC52-06NA25396.

\end{document}